\documentclass[preprint,12pt]{elsarticle}

\usepackage{amssymb,amsmath}

\usepackage{multirow}
\usepackage{graphicx}
\usepackage[ruled]{algorithm2e}

\journal{Arxiv}

\begin{document}

\begin{frontmatter}


\tnotetext[label1]{The research work of this paper were supported by the National Natural Science Foundation of China (No. 62177022, 61901165, 61501199), Collaborative Innovation Center for Informatization and Balanced Development of K-12 Education by MOE and Hubei Province (No. xtzd2021-005), and Self-determined Research Funds of CCNU from the Colleges’ Basic Research and Operation of MOE (No. CCNU22QN013).}

\author[HBUT]{Chunyan Zeng}
\author[HBUT]{Shixiong Feng}

\affiliation[HBUT]{organization={Hubei Key Laboratory for High-efficiency Utilization of Solar Energy and Operation Control of Energy Storage System, Hubei University of Technology},
            city={Wuhan},
            postcode={430000}, 
            country={China}}

\author[CCNU]{Zhifeng Wang\corref{cor1}}
\cortext[cor1]{Corresponding author}
\ead{zfwang@ccnu.edu.cn}

\affiliation[CCNU]{organization={Department of Digital Media Technology, Central China Normal University},
            city={Wuhan},
            postcode={430000}, 
            country={China}}

\author[HBUT]{Xiangkui Wan}
\author[HBUT]{Yunfan Chen}
\author[HBUT]{Nan Zhao}

\title{Spatio-Temporal Representation Learning Enhanced Source Cell-phone Recognition from Speech Recordings}

\begin{abstract}
The existing source cell-phone recognition method lacks the long-term feature characterization of the source device, resulting in inaccurate representation of the source cell-phone related features which leads to insufficient recognition accuracy. In this paper, we propose a source cell-phone recognition method based on spatio-temporal representation learning, which includes two main parts: extraction of sequential Gaussian mean matrix features and construction of a recognition model based on spatio-temporal representation learning. In the feature extraction part, based on the analysis of time-series representation of recording source signals, we extract sequential Gaussian mean matrix with long-term and short-term representation ability by using the sensitivity of Gaussian mixture model to data distribution. In the model construction part, we design a structured spatio-temporal representation learning network C3D-BiLSTM to fully characterize the spatio-temporal information, combine 3D convolutional network and bidirectional long short-term memory network for short-term spectral information and long-time fluctuation information representation learning, and achieve accurate recognition of cell-phones by fusing spatio-temporal feature information of recording source signals. The method achieves an average accuracy of 99.03\% for the closed-set recognition of 45 cell-phones under the CCNU\_Mobile dataset, and 98.18\% in small sample size experiments, with recognition performance better than the existing state-of-the-art methods. The experimental results show that the method exhibits excellent recognition performance in multi-class cell-phones recognition.
\end{abstract}



\begin{keyword}
Source cell-phone recognition \sep Temporal Gaussian feature \sep Feature representation learning  \sep 3D CNN
\end{keyword}

\end{frontmatter}


\section{Introduction}
\label{sec:introduction}
In recent decades, with the rapid development and progress of the electronic information industry, mobile electronic devices such as smartphones and tablet computers have become widely popular, and the convenience of recording, storage and transmission of digital audio has been greatly improved. The resulting phenomenon of casual recording and dissemination of digital audio has aroused social concern about digital audio security issues \cite{hanilci2011recognition}. In the passive forensic scenario of digital audio sources, the recording cell-phone models show a trend of mobility and diversification  \cite{zeng2021spatial}. The research on source cell-phone recognition identifies the source cell-phones of recordings by analyzing the differentiated feature information of different devices in audio, which can be used for evidence traceability in judicial forensics, identity authentication in multi-device collaboration scenarios and noise separation in digital audio processing.

The current mainstream methods for recording cell-phone recognition generally extract short-term spectral class features in audio as input first, and then analyze the feature information using specific classification algorithms for recognition. According to the classification algorithms used, existing mobile audio recorder recognition methods are mainly divided into two categories, namely classical methods and deep learning methods. Classical methods generally apply various types of supervised machine learning algorithms as classification decision models to distinguish cell-phone models by mapping the input feature data into higher dimensional spaces. Such methods generally require complex feature engineering steps, and the input feature data are often generic acoustic features in the field of audio recognition and are not specifically designed for source cell-phone recognition tasks; they can also be applied to tasks such as speaker recognition or sentiment recognition, and lack specificity to the extent that their efficiency and accuracy are limited. In recent years, deep learning techniques have been applied by many researchers to the field of research on recording cell-phone recognition with positive results. Deep representation learning has shown powerful capabilities in feature extraction and recognition model construction for recording cell-phone recognition, which can automatically extract highly abstract and complex feature information from raw data. However, most deep learning-based methods use generic models that are plainly data-driven, lacking domain-specific a priori information and specific structural designs for recording source signal characterization.

Comprehensive analysis of the existing work, the model level of research progress rapidly, but the feature level of research is relatively slow. Feature is the basis of source cell-phone recognition research, which determines the efficiency of information representation in audio-feature stage, and is a key issue in this research field. In addition, the source information of cell-phone in audio data contains spatial information of short-term attributes and time-varying information of long-term attributes, which makes difficult for the existing methods to characterize the long-term and short-term attributes simultaneously.

The main objective of this paper is to design a mobile audio recorder recognition method to address the existing deficiencies in information representation and recognition accuracy. We design a feature that can characterize both long-term and short-term information, and establish a corresponding spatio-temporal representation learning model with strong fitting ability of deep learning, which provides an in-depth analysis of the spatio-temporal complexity of the recording source signal simultaneously.

Our contributions in this work are:
\begin{itemize}
\item A spatio-temporal feature named sequential Gaussian mean matrix (SGMM), which is extracted from short time-series segmented acoustic data, is proposed for short-term characterization of source cell-phone.
\item We design a deep neural network named C3D-BiLSTM for source cell-phone recognition, and it has spatio-temporal representation learning capability using 3D convolutional networks (C3D) and bidirectional long and short-term memory (BiLSTM) networks, which models the long-term dependences of SGMMs.
\item We explore the effective settings of each stage through experiments, compare the fit of various approaches for this research, and analyze the characteristics of proposed method, which outperforms the stata-of-the-art methods.
\end{itemize}  

The rest of the paper is organized as follows. Section 2 presents the literature in the field including features and recognition models, and technical discussion of design solutions. Section 3 presents the design and framework of the method we proposed. Section 4 presents the experimental setups and results discussion. The last section concludes the paper with future direction.

\section{Related work}
\label{sec:related work}
In the field of cell-phone recognition research, many researchers have proposed several types of effective research methods from different perspectives. First of all, the development of acoustic features plays an extremely important role in the development of the field of cell-phone recognition, and the initial cell-phone recognition methods are usually based on the research of acoustic features, using generic acoustic features for source cell-phone recognition. As the research progressed, some researchers designed features specifically for cell-phone related information characterization based on generic acoustic features, the main characteristics of which are shown in Table 1. The development of different features in terms of connections, differences, and application scenarios are also summarized in Section 2.1.

Furthermore, in terms of cell-phone recognition models, we divide the existing recognition models into two main categories according to the classification algorithms applied: cell-phone recognition methods based on classical algorithms and cell-phone recognition methods based on deep learning techniques, which are sorted out in sections 2.2 and 2.3, respectively.

\begin{table}[h]
\footnotesize
\centering
\caption{Typical features for cell-phone recognition}
\label{tab:features}
\begin{tabular}{lll}
\hline
Feature type                               & Feature             & Main characteristics          \\ \hline
Generic acoustic features & MFCC\cite{hanilci2011recognition}           & Complex sound composition     \\
                                          & MFCC(non-speech)\cite{hanilcci2014source}   & Exogenous noise exists        \\
                                          & BED\cite{luo2018band}                       & Unilateral properties only    \\
Dedicated features        & GSV\cite{kotropoulos2014mobile}             & Temporal structure lost       \\
                                          & SRC features\cite{zou2017source}            & Lack of common representation \\
                                          & Bottleneck features\cite{li1,li2}           & Purely data-driven            \\ \hline
\end{tabular}
\end{table}

\subsection{Acoustic feature based cell-phone related information characterization}
In the past decade, a variety of acoustic features have been applied to characterize cell-phone related information in audio signals, which can be broadly classified into three categories according to the signal analysis methods applied: Cepstral class features, spectral class features, and other features of physical meaning.

Cepstral class features are widely used in the field of cell-phone recognition, and Kraetzer et al.  \cite{kraetzer2007digital} proposed using Mel cepstral features as device fingerprints to identify source recording devices, initiating the field of recording device recognition research. Based on this, C. Hanili et al. \cite{hanilci2011recognition, hanilcci2014source} used mel frequency cepstrum coefficient (MFCC) for cell-phone recognition and further extracted MFCC from silent segments in audio files, verified that features extracted from non-speech segments of audio have higher information density and recognition rate than features extracted from part of speech or whole utterance. 

In terms of spectral class features, Kotropoulos Constantine \cite{kotropoulos2014mobile} proposed the research idea of frequency domain features for feature mapping. Luo Da et al. \cite{luo2018band} used band energy difference (BED) as a feature description, and they demonstrated that the frequency response curve extracted from sample recordings can be used as an effective feature in the recognition task of source cell-phone with significant discriminative ability.

Several new ideas have also been proposed for cell-phone related features. B. Gianmarco et al.  \cite{baldini2020evaluation} evaluated the application of different entropy metrics and their applicability to microphone classification, showing that specific entropy features can provide high recognition accuracy in cell-phone recognition.

These acoustic feature-based extraction methods provide a solid foundation for accurate cell-phone recognition. However, these acoustic features can also be used in fields such as speaker recognition and emotion recognition, and more targeted feature extraction methods are lacking for the cell-phone recognition scenario.

\subsection{Classical algorithm based source cell-phone recognition methods}
Classical algorithm based source cell-phone recognition methods, generally use the extracted acoustic features such as MFCC and BED as the input of various supervised machine learning algorithms to classify cell-phones. According to the classification algorithms used, there are three main categories, support vector machines (SVM) based source cell-phone recognition model, Gaussian mixture model (GMM) based source cell-phone recognition model, and sparse representation classifier (SRC) based source cell-phone recognition model.

SVM performs nonlinear classification by dividing the data into the mapping space and uses a learning strategy of maximizing the inter-class interval. C. Hanili et al. \cite{hanilci2011recognition} used MFCC as input features, vector quantization (VQ) algorithm and SVM as back-end classification decision and analyzed the variability of spectral information of different brands of cell-phones. The effectiveness of SVM as a recognition algorithm in the source cell-phone recognition task was verified. Since then SVM has been widely used as a benchmark classification model.

The literature \cite{garcia2010automatic, eskidere2014source} uses GMM to classify cell-phones by probability density representation. The literature \cite{kotropoulos2014source} used GMM to extract more expressive cell-phone related features and input features such as MFCC and LFCC to GMM to construct Gaussian super vector (GSV), which achieved excellent recognition accuracy on classification algorithms such as SVM and radial basis neural network (RBF-NN), respectively. Jiang Y et al. \cite{jiang2019source} made some improvements on GSV feature extraction by separating the excitation signal and speech signal generated by the source cell-phone into different dimensions, and the extracted mapped GSV performs better than the original GSV in source cell-phone recognition.

SRC-based models for source cell-phone recognition perform classification by constructing complete function dictionaries and sparse representation feature data matrices.L. Zou et al. \cite{zou2017source} used GSV and MFCC to construct supervised learning dictionaries to compare and calculate the distance difference between sparse matrices to find the appropriate sample attributes for classification.

These methods use classical supervised machine learning algorithms as back-end classification decisions to distinguish categories by mapping the input feature data into higher dimensional spaces, with limited analytical mining capabilities of the data and lacking the ability to characterize the data deeply in source cell-phone recognition research.

\subsection{Deep learning based source cell-phone recognition methods}
In recent years, due to the rapid development of deep learning techniques, some researchers have applied the methods related to deep learning to the source cell-phone recognition field and achieved positive results. Part of these use deep learning techniques to automatically mine source cell-phone related information in acoustic features to extract deep features, and more researchers focus on constructing recognition network models for back-end source cell-phone recognition.

Y. Li et al. \cite{li1,li2} proposed two approaches for extracting deep features through neural networks: the first uses MFCC features to train Deep Neural Networks (DNNs) and then extracts the output of the DNN intermediate layer as features; the second uses MFCC features to train deep autoencoder networks and then uses the output of the bottleneck layer as output features. V. Vinay et al. \cite{verma2021speaker} used Discrete Fourier Transform (DFT) frequency features on a designed CNN model to compare with BED features. X. Lin et al. \cite{lin2020subband} proposed a subband-aware self-attentive mechanism based CNN to focus on the most relevant parts of the frequency band for more efficient feature representation. X. Shen et al.  \cite{shen2020rars} proposed a Residual Network (ResNet) based model for source cell-phone recognition by forming a network model by tandemly connecting ResNet and GoogLeNet. k. MichelIn et al. \cite{kulhandjian2019digital} innovatively applied blind deconvolution methods to the field of source cell-phone recognition. In \cite{zeng2020end}, we proposed a multi-feature fusion source cell-phone recognition method based on an attention mechanism. In \cite{zeng2020deep}, we use deep representation learning to extract key information features for source cell-phone recognition and propose a deep and shallow feature fusion model based on the convolutional attention mechanism.

Deep learning-based source cell-phone recognition methods have powerful representation learning capabilities to automatically extract highly abstract and complex features from raw data. However, to a certain extent, most existing recognition methods use purely data-driven generic models that do not make full use of domain a priori information and lack specific structural designs for recording source signal characterization.

Based on the previous work of our group, we found that due to the time-varying fluctuations of the audio signal amplitude and energy, the source cell-phone related information in the audio signal changes accordingly, and these changes show strong correlation in the time series analysis and are an important part of the source cell-phone related information in the audio signal. In the literature  \cite{zeng2021spatial}, we propose a parallel neural network structure in which GSV and MFCC features are input to the spatial and the temporal information extraction network, respectively, where spatial information refers to the information with short-term stability in the recording source signal, which is input to the spatial network in the form of a two-dimensional matrix tensor, and temporal information refers to the information with long-term volatility in the recording source signal. We also use a self-attentive mechanism to adaptively assign the weights of spatial and temporal information to further extract more effective information from the audio data.

In the literature \cite{zeng2021spatial}, we used two features to represent spatio-temporal information separately, but the representation of the two features on temporal and spatial information is not synchronized, which has redundancy in information representation and increases the overhead of model training. Based on the above background, the SGMM feature extraction approach proposed in this paper can characterize the spatio-temporal information in the recording source signal synchronously, and the source cell-phone related information is more fully characterized; the designed C3D-BiLSTM network mines the cell-phone related feature information on spatial scale and bidirectional time scale, which can more fully utilize the feature information and achieve accurate recognition of source cell-phone.

\section{Proposed framework}
\begin{figure}[h]
	\centering
	\includegraphics[width=1\textwidth]{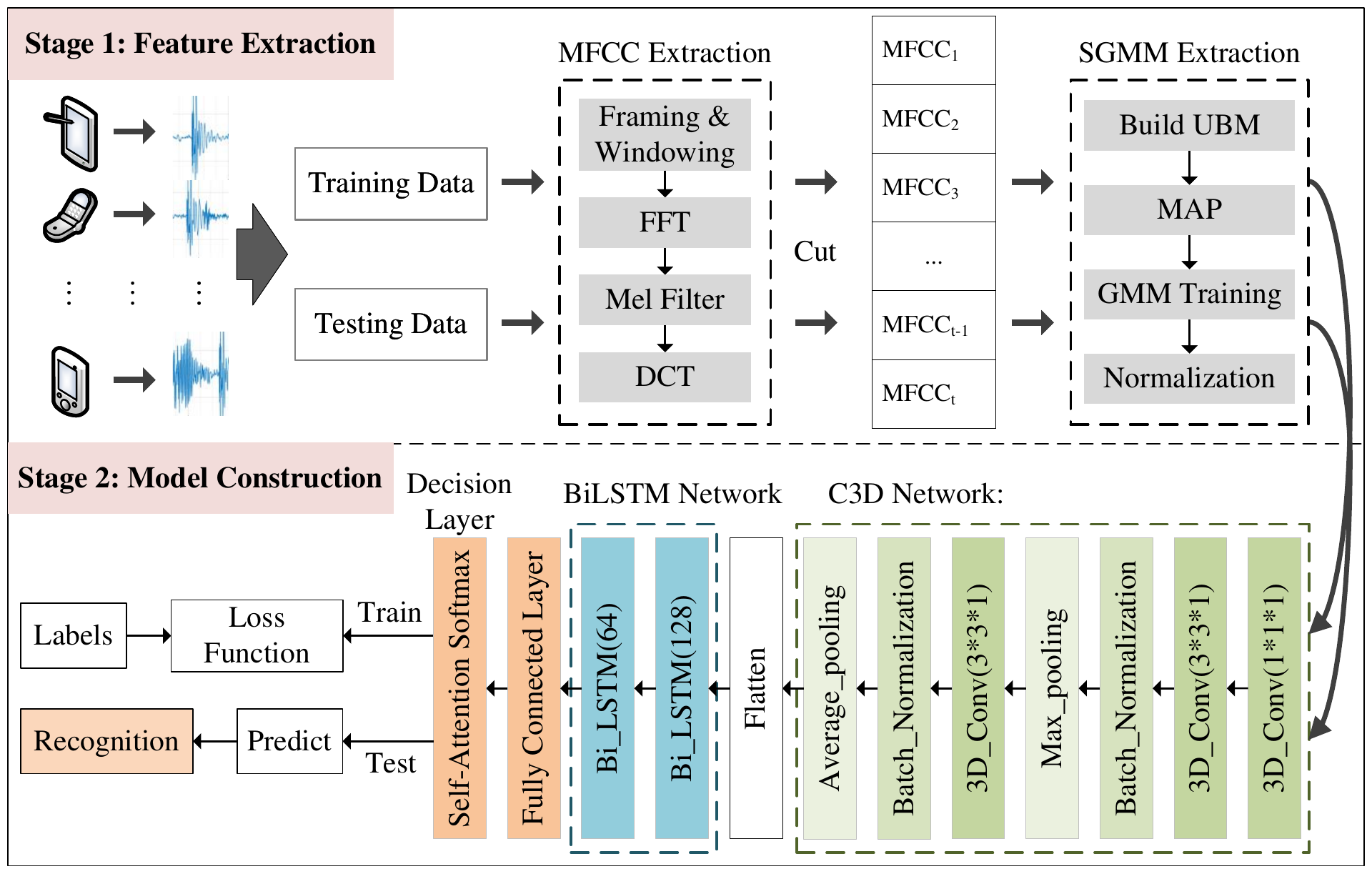}
	\caption{Framework of source cell-phone recognition based on C3D-BiLSTM spatio-temporal representation learning}
	\label{fig:framework}
\end{figure}

Our proposed source cell-phone recognition framework is shown in Figure 1, which consists of two main parts: the feature extraction stage and the model recognition construction stage, where the feature extraction stage includes the extraction process and principles of audio acoustic features MFCC and source cell-phone related features SGMM, and the model recognition stage includes the structured construction process of the spatio-temporal representation model C3D-BiLSTM and the related principles of each part of the network. . The specific processes of the key steps will be described in detail in the following subsections.

\subsection{SGMM feature extraction}
Speech signals can generally be viewed as quasi-steady state signals, namely as steady state in short term intervals and non-steady state in long term intervals. We divide the continuous speech signal into frames for short-term frequency domain cepstrum analysis, cut into temporal feature segments for long-term consideration, and use a temporal feature representation learning model to further extract the temporal information embedded in the temporal feature matrix.
In this paper, SGMM features are extracted from the temporal feature segments, and the extracted SGMM features can efficiently characterize the time-domain and frequency-domain information of the recording source signal.SGMM features are feature matrices extracted using Gaussian mixture models, which can accurately represent the attribute features of the data by probability density models.The extraction steps of SGMM features are shown in Algorithm 1.

\begin{algorithm}[h]
\caption{SGMM feature extraction process}
\LinesNumbered
\KwIn{MFCCs}
\KwOut{SGMM feature: GMM.$\mu$}
    Initialization: $\omega$ :weights, $\mu$ :means, $\sigma$ :covariances\;
    Reshape MFCC into time-series feature matrix in $M$*$T$ format\;
    \qquad $M$: the dimensions of MFCC\\
    \qquad $T$: the frames number of the each short-term period\\
    Apply EM algorithm to update parameters: \\
    \qquad E-Step: Calculate the conditional probability expectation of the joint distribution: \\
    \hspace{4em} ${Q_i}({z_i}) = p({z_i}|{x_i},{q_j})$\\
    \hspace{4em} $l(q,{q_j}) = \sum\limits_{i = 1}^n {\sum\limits_{{z_i}} {{Q_i}({z_i})log\frac{{p({x_i},{z_i};q)}}{{{Q_i}({z_i})}}} } $ \\
    \qquad M-Step: Maximize \ $l(\theta ,{\theta _j})$ \ and get \ ${\theta _{j + 1}}$ \ : \\
    \hspace{4em} ${\theta _{j + 1}} = \arg \max (l(\theta ,{\theta _j}))$ \\
    Established the UBM= $\{ {\omega _{UBM}},{\mu _{UBM}},{\sigma _{UBM}}\} $\;
\For{n=1:cell-phones}{
    \For{s=1:samples}{
    MAP-adapts: fits a nmix-component GMM to data\;
    Min-Max scaling normalization\;
    }
}
Obtain a structure containing the GMM hyperparameters of data\;
\textbf{Return}  GMM.$\omega$, GMM.$\mu$, GMM.$\sigma$
\end{algorithm}

\subsubsection{Acoustic feature - short-term frequency spectrum characterization}
MFCC feature is a frequency cepstrum feature based on short-term Fourier transform, which is one of the most commonly used features in the field of recognition of source cell-phone and can well characterize acoustic feature information. the extraction process of MFCC includes pre-processing (frame splitting, windowing), Fast Fourier Transform (FFT), Mel filtering, logarithmic operation, Discrete Cosine Transform (DCT) and other steps. The specific extraction process is as follows: 
\begin{enumerate}[1)]
\item First, in order to get a stable representation of the speech signal in the frequency domain, the whole speech signal s(n) needs to be framed. The framing of the audio signal is achieved by using a shiftable fixed-length window for weighting, namely, the input speech is framed and windowed, the frame length is taken as $fl$, a part of $fl$ is taken as the frame shift $fs$, and the window is added using the Hamming window function; 
\item Then obtains the spectral information of each frame by performing a fast Fourier transform on the framed and windowed signal;
\item Uses the obtained amplitude spectrum signal F(f) to be filtered through a Mel filter bank, and filters the amplitude spectrum signal F(f) through a set of Mel-scale triangular filter banks whose center frequencies are uniformly distributed on the Mel scale; 
\item Then calculates each filter The logarithmic amplitude spectrum at the output of each filter group is then calculated by DCT to obtain the MFCC features.
\end{enumerate}  

We experimentally compare the recognition accuracy of cell-phones in different frequency bands and different frequency band coverage in the MFCC extraction process, and determine the frequency band range where the source cell-phone related information in the audio signal is located. And through frame processing experiments, we derived the frame length and frame shift parameter settings that match the cell-phone recognition characteristics.

\subsubsection{Temporal Gaussian feature - source cell-phone related information characterization}
The audio signal is mixed with a variety of noises from the sound source to the process of audio file generation, and its composition is shown schematically in Figure 2, where the composition of the audio signal $a(t)$ is given by the formula:
\begin{equation}
    a(t) = \left\lbrack s(t) + n(t) \right\rbrack*d(t)
\end{equation}
Where $s(t)$ denotes the human voice in the source part, $n(t)$ denotes the ambient noise, $d(t)$ denotes the device floor noise, and $*$ denotes the convolution calculation. 

\begin{figure}[h]
	\centering
	\includegraphics{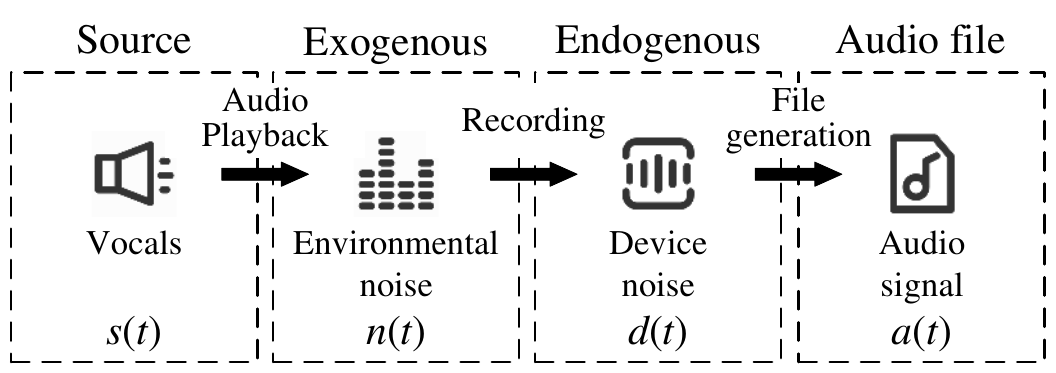}
	\caption{Components of the audio signal}
	\label{fig:audio}
\end{figure}

The digital audio file data obtained by recording carries a variety of information, such as speech content, speaker information, environmental noise, etc., which also contains information associated with the recording device, so the directly extracted acoustic features will contain a variety of interfering information. In such related tasks as speech content recognition or speaker recognition, the feature information used is the biometric features of the speaker's voice, which are more clearly expressed in speech and can be distinguished as distinctly different with the human ear.

In contrast, the characteristic information used in the audio source recognition task of recording devices comes from the differential characteristics generated during the transmission of digital signals, which is a noise-like presence that appears in a specific frequency band. This feature information comes from the differences in the internal circuit structure and components of different recording cell-phones, caused by the variability in the transmission of the speech excitation signals generated during the speech recording process in the device circuitry \cite{hanilci2011recognition}, and this variability causes the recording cell-phones to leave traces of distinguishing features in the audio files during the speech capture process, and we use a Gaussian mixture model to mine this cell-phone in the audio feature information and extract the source cell-phone related features with higher information density from the acoustic features.

\begin{figure}[h]
	\centering
	\includegraphics{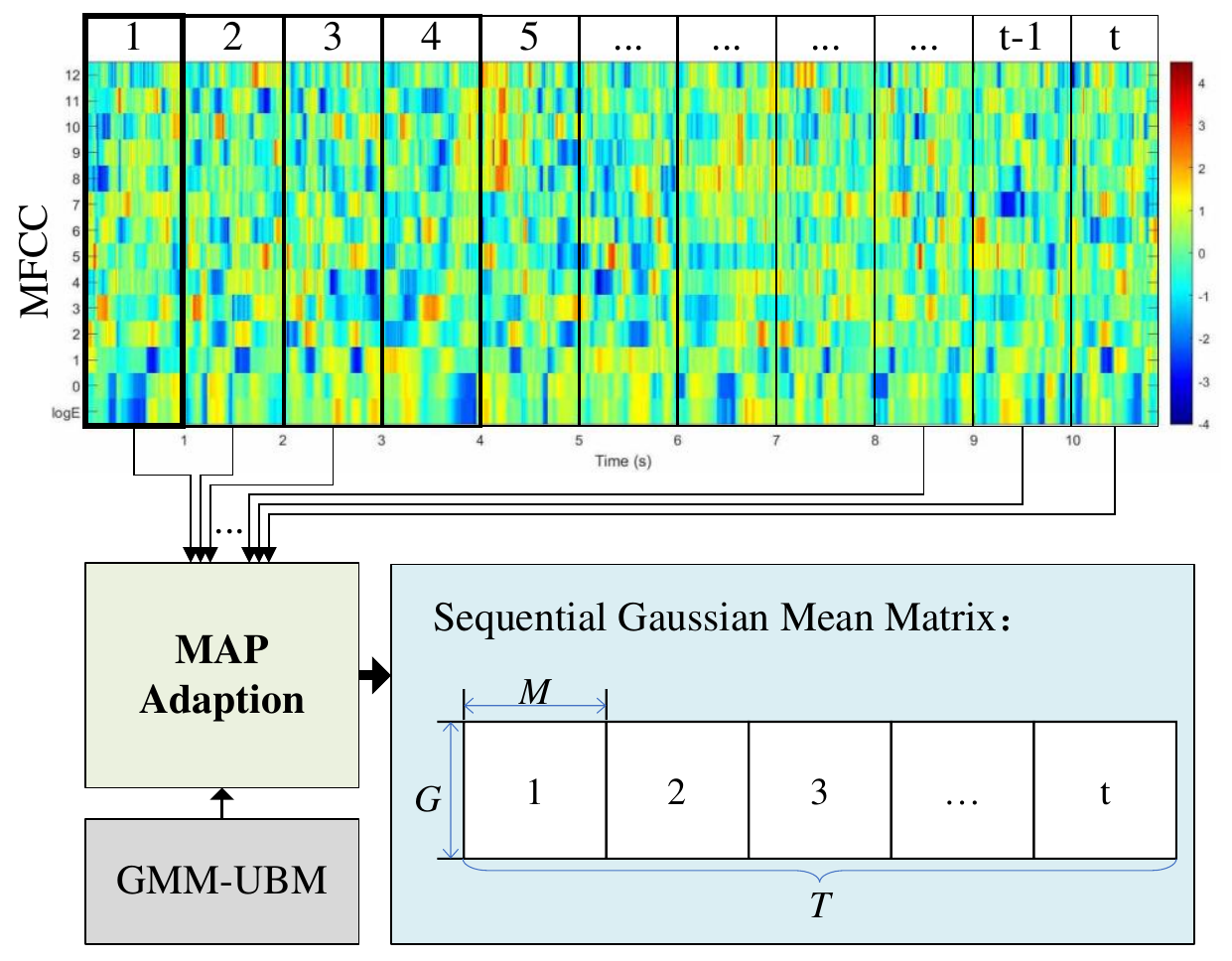}
	\caption{SGMM feature extraction process}
	\label{fig:SGMM}
\end{figure}

Gaussian supervector class features have been shown to be effective in the research of source recognition of cell-phones \cite{zeng2021spatial, kotropoulos2014mobile, jiang2019source}, and the core idea is that the probability distribution of any shape can be approximated by multiple Gaussian distribution functions.The SGMM feature extraction process is shown in Figure 3, and the steps of extracting Gaussian supervector class features by GMM are as follows: firstly, the GMM is trained by a large amount of speech data, and it is trained by the expectation-maximization ( Expectation-Maximum, EM) algorithm for iterative training to obtain a diagonal covariance universal background model (UBM) with G Gaussian mixture components, where the EM algorithm is divided into two steps, and the first step calculates the conditional probability expectation of the joint distribution:
\begin{equation}
    l\left( {q,q_{j}} \right) = {\sum_{i = 1}^{n}{\sum_{z_{i}}{Q_{i}\left( z_{i} \right)log\frac{p\left( {x_{i},z_{i};q} \right)}{Q_{i}\left( z_{i} \right)}}}}
\end{equation}
where the probability $Q_i$ is calculated by the following equation:
\begin{equation}
    Q_{i}\left( z_{i} \right) = p\left( z_{i} \middle| {x_{i},q_{j}} \right)
\end{equation}
The second step iteratively updates to maximize the expectation value:
\begin{equation}
    \left. \theta_{j + 1} = {\mathit{argmax}(}l\left( {\theta,\theta_{j}} \right) \right)
\end{equation}
The weight coefficient matrix, mean coefficient matrix and covariance coefficient matrix with G number of Gaussian components are obtained. Then, by using the maximum a posteriori (MAP) algorithm, the GMM of the target device is obtained adaptively for the feature vector of each device in the training data, and the GMM mean coefficient matrix of the input data is obtained by calculation, and after the maximum-minimum normalization, that is, the SGMM features of the form $M\times G\times T$ are obtained, where $M$ is the the number of feature dimensions obtained by MFCC extraction, $G$ is the number of mixed Gaussian components in the SGMM extraction process, and $T$ denotes the sequence length of the temporal feature matrix.

In the general process of extracting Gaussian supervector class features based on Gaussian mixture model, the feature information carried in the input features is streamlined and refined, and the information density and utilization efficiency are enhanced and improved, but simultaneously the temporal information carried by the input features is also lost. Here we use MFCC as input features, and the extracted MFCC features are divided into one feature matrix segment per several consecutive frames and input to GMM for Gaussian mean matrix feature extraction. $T$ segments of SGMM feature matrix segments ($t$ frames/segments) are extracted from all MFCC feature frames obtained from each speech sample, preserving the temporal structure of the feature matrix to enable the embedded long-time fluctuation characteristic information not to be lost.

\subsection{C3D-BiLSTM spatio-temporal representation learning model construction}
The main network structure of the proposed spatio-temporal feature representation learning model for source cell-phone recognition consists of a spatial information representation learning network and a temporal information representation learning network, and the block diagram is shown in Figure 3. The spatial information feature representation learning network and the temporal information feature representation learning network use 3D convolutional network and BiLSTM network, respectively, where the temporal network functions after the spatial network and the output of the temporal information feature representation learning network is used as the output of the final connection to the decision layer, and the network structure of this model is described in detail below.

\begin{figure}[h]
	\centering
	\includegraphics[width=1\textwidth]{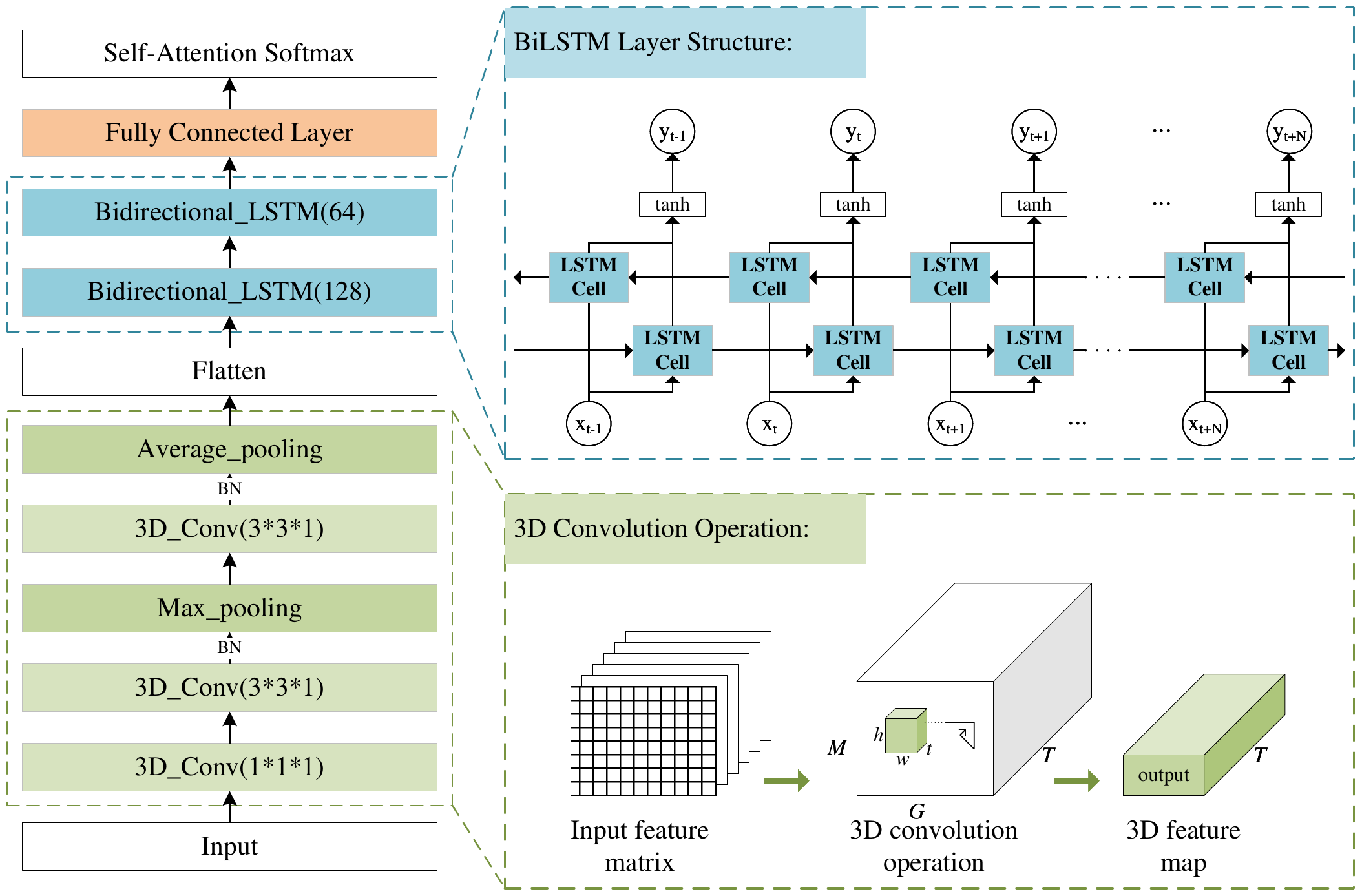}
	\caption{C3D-BiLSTM spatio-temporal representation learning model}
	\label{fig:C3D-BiLSTM}
\end{figure}

\subsubsection{Spatial information feature representation learning network}
In the front-end part of the spatio-temporal feature representation learning model, we construct a 3D convolutional network to learn the structured representation of spatial information on the input temporal feature matrix data. 3D convolutional network has one more dimensional depth information in its input feature map than the $H\times W$ of 2D convolution, and its convolutional kernel moves in three directions, then there exist three degree-of-freedoms in the motion of the convolutional kernel. For the feature matrix of the temporal structure, its input dimension is $H\times W\times T$, where T denotes the sequence length of the temporal feature matrix. In the process of using 3D convolutional network to process SGMM features, the representation learning is mainly aimed at the spatial information of the features, and we use the dimension about the depth of the features in 3D convolution as the time scale to be able to preserve the temporal structure of the input feature matrix in the convolution operation to subsequently make full use of the fluctuation information of the time dimension.

The construction of 3D convolutional networks follows the two main properties of local connectivity and weight sharing of convolutional neural networks. In this structure, when convolution operations are performed on the feature matrices of multiple consecutive time series, the feature values computed between layers are connected to the feature matrices of multiple neighboring time points in the previous layer to capture the time-varying fluctuation information of the features. Meanwhile the weight sharing mechanism can greatly reduce the number of parameters in the network and improve the computational efficiency of feature extraction in the temporal space. The convolution operation equation is :
\begin{equation}
    out = \sigma\left( {W*X + b} \right)
\end{equation}
where $X$ is the input value, $W$ denotes the value of the convolution kernel, $b$ is the bias value, $*$ denotes the convolution operation, and $\sigma$ denotes the activation function operation.

In addition, we add a 1 × 1 convolutional kernel after the input layer for nonlinear mapping, which allows information transfer with fewer connections through low-dimensional embeddings, thus producing sparser features for efficient computation. The obtained feature maps are then pooled to reduce the number of parameters, using maximum pooling in the middle layer of the network to retain the valid information in the features and average pooling in the post-network layer to reduce information loss.

We use a 3D batch normalization layer \cite{ioffe2015batch} between the convolutional layers, which can handle a four-dimensional tensor, and normalize the mean and variance by channel based on the current batch data to speed up the training and improve the model accuracy. For training, $\mu_\beta$ and $\sigma_\beta^2$ are the statistics of the current batch, which represent the mean and variance, respectively, and are calculated as:
\begin{equation}
    \mu_{\beta} = \frac{1}{m}{\sum_{i = 1}^{m}x_{i}}
\end{equation}
\begin{equation}
    \sigma_{\beta}^{2} = \frac{1}{m}{\sum_{i = 1}^{m}\left( {x_{i} - \mu_{\beta}} \right)^{2}}
\end{equation}
The normalized function equation is given by:
\begin{equation}
    \overset{\hat{{x_{i}}}} = \frac{x_{i} - \mu_{\beta}}{\sqrt{\sigma_{\beta}^{2} + \epsilon}}
\end{equation}
\begin{equation}
    y_{i} = \gamma\overset{\hat{{x_{i}}}} + \beta
\end{equation}
where $\gamma$ and $\beta$ are the learnable parameters in the BN minibatch, referring to the proportion and offset, respectively.

\subsubsection{Temporal information feature representation learning network}
In the temporal information feature representation learning network part, we construct a BiLSTM network to perform temporal analysis on the output of the C3D network. the BiLSTM network uses the last layer of features of the 3D convolutional network as input, which has a recursive feedback structure connected by chains to further extract temporal feature information from the data.

BiLSTM is a combination of forward LSTM and backward LSTM, and compared with unidirectional LSTM, BiLSTM is able to analyze and consider the forward and backward time series information synchronously when dealing with sequential time series information, so as to better grasp the information correlation in both directions \cite{alom2019state}.BiLSTM can deal with long-term volatility by analyzing the time series feature matrix moment by moment , removing or adding information operations to the temporal state of each cell through the gating structure, and passing the useful feature information to the next temporal cell.

The cell structure of BiLSTM mainly controls the update and utilization of the temporal information through three gates: the input gate controls the reading of new information, the forgetting gate controls the discarding of information, and the output gate controls the output of information. Its parameter update formula is shown in equation (10-14), where ${X}_t$ is the vector input at time $t$; ${h}_{t-1}$ is the output at time $t-1$; ${C}_{t-1}$ is the cell state at time $t-1$; ${i}_t$ is the input gate; ${f}_ t$ is the forgetting gate; ${o}_t$ is the output gate; $\sigma$ is the activation function, and the default is the sigmoid nonlinear function.
\begin{equation}
    \left| \mathbf{i}_{t} \right| = \sigma_{i}\left( {\mathbf{w}_{ix}\mathbf{X}_{t} + \mathbf{w}_{ih}\mathbf{h}_{t - 1} + \mathbf{w}_{ic}\mathbf{C}_{t - 1} + b_{i}} \right)
\end{equation}
\begin{equation}
    \left| \mathbf{f}_{t} \right| = \sigma_{f}\left( {\mathbf{w}_{fx}\mathbf{X}_{t} + \mathbf{w}_{fh}\mathbf{h}_{t - 1} + \mathbf{w}_{fc}\mathbf{C}_{t - 1} + b_{f}} \right)
\end{equation}
\begin{equation}
    \left| \mathbf{o}_{t} \right| = \sigma_{o}\left( {\mathbf{w}_{ox}\mathbf{X}_{t} + \mathbf{w}_{oh}\mathbf{h}_{t - 1} + \mathbf{w}_{oc}\mathbf{C}_{t} + b_{o}} \right)
\end{equation}
\begin{equation}
    \left| \mathbf{C}_{t} \right| = \mathbf{f}_{t}\mathbf{C}_{t - 1} + \mathbf{i}_{t}{\tanh\left( {\mathbf{w}_{cx}\mathbf{X}_{t} + \mathbf{w}_{ch}\mathbf{h}_{t - 1} + b_{c}} \right)}
\end{equation}
\begin{equation}
    \left| \mathbf{h}_{t} \right| = \mathbf{o}_{t}{\tanh\left( {\left| \mathbf{C}_{t} \right| + b} \right)}
\end{equation}
where: ${w}_{ix}$, ${w}_{fx}$, ${w}_{ox}$, ${w}_{cx}$ are the weights of the input gate, forgetting gate, output gate and cell state on the current input ${X}_t$, respectively; ${w}_{ih}$, ${w}_{fh}$, ${w}_{oh}$ and ${w}_{ch}$ are the weights of the input gate, forgetting gate, output gate and cell state on the output ${h}_t$, respectively; ${w}_{ic}$, ${w}_{fc}$ and $ {w}_{oc}$ are the weights of input gate, forgetting gate and output gate on the cell state ${C}_t$, respectively; $b$ is the bias of each gate.

\subsubsection{Back-end classification of spatio-temporal representation learning model}
We use cross entropy as the loss function of the spatio-temporal feature representation learning model for source cell-phone recognition, and we use Adam \cite{kingma2014adam} in the spatio-temporal representation model to achieve optimization of the network parameters \cite{alom2019state}, and the detailed optimization process is shown in Algorithm 2.

\begin{algorithm}[h]
\caption{Optimization algorithm for spatio-temporal representation learning model parameters}
\LinesNumbered
\KwIn{learning rate $ \alpha$, exponential decay rate {$ \beta_1$}and {$ \beta_2$} , $ \epsilon$}
\KwOut{weights $ W $}
Initialization: $ W $, $m$, $v$\\
\For{t=1:T(T=samples/batch size)}{
    Compute exponential moving average of the gradient at each moment\;
    \qquad ${m^t} = {\beta _1}{m^{t - 1}} + (1 - {\beta _1})dW^{t-1}$\\
    Compute exponential moving average of the gradient square at each moment\;
    \qquad ${v^t} = {\beta _2}{v^{t - 1}} + (1 - {\beta _2})dW^{t-1}$\\
    Bias correction\;
    \qquad ${\hat m^t} = {m^t}/(1 - \beta _1^t)$\\
    \qquad ${\hat v^t} = {v^t}/(1 - \beta _2^t)$\\
    Update weights\;
    \qquad ${W^t} = {W^{t - 1}} - \alpha (\frac{{{{\hat m}^t}}}{{\sqrt {{{\hat v}^t}}  + \epsilon }})$
}
\end{algorithm}

The back-end decisions use a Softmax classifier with a fully connected layer connected to a representation learning network that maps the learned distributed feature representations to the sample labeling space. The algorithm for learning the objective function to loss values and weights is shown in Algorithm 3.

\subsection*{C. Back-end classification for source cell-phone recognition}

\begin{algorithm}[h]
\caption{Algorithm for model objective function learning}
\LinesNumbered
\KwIn{BiLSTM layers output $o$, learning rate $ \alpha$}
Initialization: weights $ W $, cross-entropy loss $ C$=0\\
\For{t=1:T(T=epochs)}{
    Compute output $S^t$ of softmax layer\;
    \qquad ${S_i} = \frac{{{e^{o_i}}}}{{\sum\nolimits_j {{e^{o_j}}} }}$\\
    Compute cross-entropy Loss $L^t$ \;
    \qquad $C =  - \sum\nolimits_i^n {{y_i}\log ({S_i})}$\\
    Backpropagation error\;
    \qquad $X=W*o$\\
    \qquad $\frac{{\partial C}}{{\partial W}} = \frac{{\partial C}}{{\partial X}}*\frac{{\partial X}}{{\partial W}} = \frac{{\partial C}}{{\partial X}}*o$\\
    Update weights\;
    \qquad $W^t = W^{t - 1} - \alpha (\frac{{\partial {C^{t - 1}}}}{{\partial W^{t - 1}}})$\\
}
\end{algorithm}

In order to make the network model learn more discriminative intrinsic feature information during the training process, we use a self-attention mechanism in the back-end decision-making for the sparse representation of the intrinsic feature information matrix. distribution to stabilize the training process, and finally obtain the distribution of the correlation or similarity between each sample and each label information. The formula of the self-attention mechanism is:
\begin{equation}
    Attention(Q,K,V) = softmax\left( \frac{Q*K^{T}}{\sqrt{d}} \right)*V
\end{equation}
where $Q$, $K$ and $V$ denote the three input parameters of the self-attentive module, which have the same dimension, and $d$ denotes the size of the last dimension of these three parameters.

\section{Experimental setups and results discussion}
All experiments in this paper were conducted on Matlab 2020a and TensorFlow 2.1. The relevant experimental hardware configurations are CPU: Intel Xeon Gold 5218×2, GPU: NVIDIA TITAN RTX (24GB video memory), and memory: 32GB. Experimental data are used from the CCNU\_Mobile dataset \cite{zeng2020end}.

\subsection{Experimental database and setup}
\begin{table}[h]
\centering
  \caption{Brands and models of cell-phone in the CCNU\_Mobile dataset \cite{zeng2020end}}
  \label{table:dataset}
  \begin{tabular}{llll}
   \hline
    Brands  & Models\\ 
    \hline
    APPLE   & iPhone6(4), iPhone6s(3),iPhone7p, iPhoneX, \\ 
            & iPhoneSE, iPad7, Air1, Air2(2)\\
    XIAOMI  & mi2s, note3, mi5, mi8, mi8se(2), mix2,\\
            & redmiNote4x, redmi3S\\
    HUAWEI  & Nova, Nova2s, Nova3e, P10, P20, TAG-AL00\\
    HONOR   & honor7x, honor8(3), honorV8, honor9, honor10\\
    VIVO    & x3f, x7, y11t\\
    ZTE     & C880a, G719c\\
    SAMSUNG & S8, sphd710\\
    OPPO    & R9s\\
    NUBIA   & Z11\\ 
    \hline
  \end{tabular}
\end{table}

The CCNU\_Mobile dataset was recorded from 45 different cell-phone models manufactured by 9 different brands, including Apple, Huawei, Honor, Nubia, oppo, vivo, Xiaomi, Samsung and ZTE, with the cell-phone models shown in Table 2. The recorded corpus is from the TIMIT dataset, with 642 audio samples recorded from each cell-phone, and we randomly select 514 samples from each cell-phone as the training set and another 128 samples as the test set.

The validation experiment process is as follows: 
\begin{enumerate}[1)]
    \item For each sample, we first perform MFCC feature extraction using the melcepst function in Matlab speech processing toolbox voicebox to obtain the feature matrix of each speech sample. 
    \item For the obtained MFCC feature matrix, we first perform temporal segmentation and then input the temporal feature matrix into the GMM to model each recording cell-phone. After training to obtain the GMM for each recording cell-phone, the Gaussian supervector is calculated by superimposing the mean matrix and the main diagonal of the covariance matrix of all components to obtain the SGMM features. The time series information of the feature matrix is preserved throughout the feature extraction process. 
    \item The SGMM features are input to the spatio-temporal representation learning model we constructed (see Section 3.2 of this paper for the specific network structure), and the softmax probability classification results are obtained through network training and optimization, and the training model is tested in the test set. The neural network model structure is implemented by Tensorflow, using Adam as the optimizer, with an initial learning rate of 0.1. Every 30 epochs, the learning rate decreases to one-tenth of the previous rate.
\end{enumerate}

We conducted multiple sets of controlled experiments on the CCNU\_Mobile dataset, using recognition accuracy on the test set as the main evaluation index, and set up multiple sets of experiments to analyze and demonstrate the effectiveness of the proposed method at multiple levels, such as features, models, and parameter settings.

\subsection{Experimental results and analysis}
\subsubsection{Verifying the effect of frequency bands in the acoustic feature extraction process}
In order to explore the distribution frequency bands of source cell-phone related information in audio signals, a set of control variable experiments is set up in this paper to compare the effect of feature extraction frequency bands on recognition results by adjusting the frequency intervals extracted by MFCC. The set of experiments is conducted on the CCNU\_Mobile dataset, containing 45 cell-phone models, with 20 speech samples taken from each model. In the process of MFCC extraction, other parameters are kept constant, and different filtering frequency thresholds are set for the MFCC extraction band range, using SVM as the classification and recognition model.

The recognition accuracy of MFCC extracted by different frequency bands on SVM is shown in Table 3. Through the analysis of the experimental results, we found that MFCC extracted by low and middle frequency bands got higher recognition accuracy, which indicates that the information related to the source cell-phone in the audio signal is mainly concentrated in the low and middle frequency bands, especially in the low frequency bands, but there is also a small distribution in the high frequency bands. Therefore, we will use the acoustic features extracted from the low and middle frequency bands as the input for the subsequent feature extraction stage.

\begin{table}[h]
\centering
\caption{Comparison of different frequency bands during feature extraction.}
\label{tab:frequency bands}
\begin{tabular}{@{}lll@{}}
  \hline
  \multirow{2}{*}{\begin{tabular}{@{}l@{}}Low End of\\ Lowest Filter(Hz)\end{tabular}} &
  \multirow{2}{*}{\begin{tabular}{@{}l@{}}High End of\\ Highest Filter(Hz)\end{tabular}} &
  \multirow{2}{*}{\begin{tabular}{@{}l@{}}Accuracy\end{tabular}} \\ \\ 
  \hline
 {0}  & 9600  & 77.40\% \\
      & 16000 & 86.90\% \\
      & 32000 & 87.30\% \\
 6400 & 16000 & 76.40\% \\
 9600 & 32000 & 66.40\% \\ 
  \hline
\end{tabular}
\end{table}

\subsubsection{Verifying the effect of frame length and frame shift during acoustic feature extraction}
Audio signal is time-varying signal, various information of audio and parameters characterizing its basic features change with time but remain basically constant in a short period of time, so audio signal has long-term volatility and short-term stability. In the step of extracting MFCC, the speech needs to be framed and windowed before FFT, and the appropriate parameter settings of frame length $fl$ and frame shift $fs$ can make the classification model capture the volatility of the audio signal more accurately.

A shorter frame length setting reflects the transient changes in the frequency characteristics of the audio signal, and is suitable for audio signals with large fluctuations. Longer frame lengths provide a better indication of the commonality of the frequency characteristics of the audio signal and are suitable for smoother audio signals. In addition, it is necessary to keep overlapping frame shifts between frames in order to make smooth transitions and maintain continuity between frames.

\begin{table}[h]
\centering
\caption{Comparison of different frame length and frame shift during MFCC extraction.}
\label{tab:frames}
\begin{tabular}{@{}lll@{}}
\hline
  \begin{tabular}[c]{@{}c@{}}$fl$ (ms)\end{tabular} &
  \begin{tabular}[c]{@{}c@{}}$fs$ (ms)\end{tabular} &
  \begin{tabular}[c]{@{}c@{}}Accuracy\end{tabular} \\ \hline 
  64  & 16  & 54.7\% \\
      & 32 & 53.1\% \\
  128 & 32  & 67.4\% \\
      & 64 & 64.7\% \\
  256 & 64  & 87.3\% \\
      & 128 & 86.2\% \\
  512 & 128 & 81.3\% \\
      & 256 & 78.9\% \\
\hline
\end{tabular}
\end{table}

In order to research the effects of frame length and frame shift on MFCC feature quality, we set up a set of comparison experiments, taking 20 speech samples from each of the 45 cell-phone models in the CCNU\_Mobile dataset as experimental data, and setting the parameters for MFCC extraction as follows: frame length of 64ms, 128ms, 256ms, 512ms, and setting the frame increments to 1 /4 and 1/2 of the frame length.

The recognition accuracy on SVM for the comparison test with the combination of different frame lengths and different frame shifts set in MFCC extraction is shown in Table 4. The experimental results show that the longer frame length and shorter frame shift settings are more suitable for representing the source cell-phone related feature information in the audio signal. The longer frame length setting can highlight the commonality of the frequency band distribution characteristics in this time interval, while the shorter frame shift setting reduces the overlap and redundancy of information on the basis of ensuring a smooth transition between frames. 

\subsubsection{Comparison of characterization performance with existing typical features}
We selected five typical features to compare and verify the ability of various types of features to characterize recording source information, including the classical acoustic features MFCC, I-vector, BED and the features GSV and SGMM that have been refined for source cell-phone related information. experiments were conducted on the CCNU\_Mobile dataset, using all samples in the set for experiments, and uniformly using the classical supervised machine learning algorithm SVM for classification recognition to compare the performance of different features in the cell-phone effect source recognition task, and the results are shown in Table 5.

\begin{table}[h]
\centering
  \caption{Comparison of different typical features.}
  \label{tab:feature test}
  \begin{tabular}{@{}ll@{}}
    \hline
    Feature       & Accuracy \\ 
    \hline
    MFCC\cite{hanilci2011recognition}         & 94.5\%  \\
    I\-vector         & 96.5\%  \\
    BED\cite{luo2018band}        & 70.9\%  \\
    GSV\cite{kotropoulos2014mobile}        & 96.74\% \\
    SGMM        & 98.78\% \\ 
    \hline
  \end{tabular}
\end{table}

As shown in Table 5, compared with classical acoustic features such as MFCC, GSV features can effectively refine the spatial information related to the source cell-phone in the data, which results in higher classification accuracy. In contrast, the classification accuracy under SGMM features is lower than that of GSV, indicating that SVM cannot effectively utilize the temporal information in SGMM features.

\subsubsection{Experiment to test the fit of various models with SGMM features}
SGMM feature is a Gaussian super-vector class feature with temporal structure, and its high characterization ability needs to be reflected by the recognition accuracy in experiments. In the process of designing recognition network models, we took into account the 3D temporal matrix structure of SGMM feature, selected network layers with correspondingly excellent characterization power of temporal feature information for design, and compared the fit of various types of networks with SGMM features.
During the model design process, considering the characteristics of the source cell-phone recognition task, we selected various classical deep neural networks for testing, and finally used the excellent-performing C3D network with BiLSTM network to build the model, and tested and optimized the network structure and parameters for several times to build a stable and effective recognition model. Since the input format of SGMM features is $M\times G\times T$, which contains three dimensions, we input SGMM features in the form of feature matrix in ($M\times G,T$) format for networks with two-dimensional input scales, and we input SGMM features in the form of three-dimensional feature matrix in (M,G,T) format for networks with three-dimensional input scales.

\begin{table}[h]
\centering
  \caption{Comparison of different deep learning methods.}
  \label{tab:deep}
  \begin{tabular}{@{}lll@{}}
    \hline
    Model          &  Loss      &  Accuracy \\
    \hline
    DNN            & 2.09       & 95.82\%  \\
    CNN            & 2.38       & 96.41\%  \\
    ResNet18       & 2.12       & 97.05\%  \\
    BiLSTM         & 3.06       & 94.60\%  \\
    C3D            & 2.36       & 98.22\% \\
    C3D-BiLSTM     & 2.23       & 99.03\% \\ 
    \hline
  \end{tabular}
\end{table}

From the experimental results in Table 6, it is shown that for the recognition of SGMM features, the convolutional class network shows a strong characterization of the feature matrix in the network with two-dimensional input scales, and ResNet has a slight increase in accuracy on top of that. In comparison with the network at the two-dimensional input scale, the network at the three-dimensional input scale performs even better. The 3D convolutional neural network fits the input form of SGMM features in three dimensions well with degrees of freedom in three directions, and the model in this paper incorporates BiLSTM in the network design to further exploit the information of temporal fluctuations in the data and shows a better task matching.

\subsubsection{Comparison of recognition performance with different existing methods}
We compare the recognition performance of our method with existing state-of-the-art methods on the CCNU\_Mobile dataset. The literature \cite{hanilci2011recognition}, as a typical approach of classical methods in the field of source cell-phone recognition, often appears as benchmark experiments in the existing literature, which uses acoustic features MFCC input to SVM classifier; in the literature   \cite{kotropoulos2014mobile}, a Gaussian mixture model is used to extract GSV and then input to SVM for classification; in the literature \cite{verma2021speaker}, a CNN network is constructed as a recognition model, and in our experiments we use GSV as feature input; in literature \cite{zeng2020end}, multiple features are spliced and input to a feature fusion network constructed by CNN and DNN for recognition, and the weights of feature fusion are assigned using an attention mechanism; in literature  \cite{zeng2021spatial}, a recognition model named parallel spatio-temporal neural network (PSTNN) is constructed, and MFCC and GSV are used as inputs.

\begin{table}[h]
\centering
  \caption{Comparison of different methods.}
  \label{tab:methods}
  \begin{tabular}{@{}llll@{}}
    \hline
    Method          &   Model   &   Feature &   Accuracy \\ 
    \hline
    MFCC-SVM\cite{hanilci2011recognition}     
                    &   SVM         &   MFCC        &   87.46\% \\
    GSV-SVM\cite{kotropoulos2014mobile}
                    &   SVM         &   GSV         &   94.48\% \\
    CNN\cite{verma2021speaker}
                    &   CNN         &   GSV         &   94.50\% \\
    Features fusion model\cite{zeng2020end} 
                    &   CNN,DNN     &   MFCC,GSV,i-vector     &   97.30\% \\
    PSTNN\cite{zeng2021spatial}      
                    &   DNN,BiLSTM  &   MFCC,GSV    &   97.50\% \\
    C3D-BiLSTM      & C3D,BiLSTM    &   SGMM        &   99.03\% \\ 
    \hline
  \end{tabular}
\end{table}

To verify the advancedness of the proposed method in this paper, we compared the recognition performance of various methods on the same dataset, using the average recognition accuracy as the evaluation index. The average accuracy rates of our method and five existing advanced methods are shown in Table 7. The experimental results show that the recognition performance of our proposed method outperforms other methods and improves the accuracy rate by 1.53\% $\sim$ 11.74\% on the benchmark experiment, which verifies the effectiveness of our proposed SGMM feature and spatio-temporal representation learning model.

\subsubsection{Recognition performance test experiment under the condition of small sample size}
In actual forensic scenarios, the situation of small samples of forensic data and short sample length is often faced. The recognition performance experiments under the condition of small sample size can test the ability of the model to quickly adapt to multi-class recognition under the limited number of samples. To test the robustness of our proposed source cell-phone recognition method under the condition of small sample size, we use only 5 samples for each cell-phone model for training, and the test results are shown in Table 8.

\begin{table}[h]
\centering
  \caption{Experiments under small sample size conditions}
  \label{tab:small sample}
  \begin{tabular}{@{}lll@{}}
    \hline
    Model  &  Feature     & Accuracy \\ 
    \hline
    SVM            & MFCC        & 82.85\%  \\
    CNN            & MFCC        & 90.12\%  \\
    CNN            & GSV         & 93.87\%  \\
    BiLSTM         & SGMM        & 90.56\%  \\
    C3D            & SGMM        & 97.27\% \\
    C3D-BiLSTM     & SGMM        & 98.18\% \\ 
    \hline
  \end{tabular}
\end{table}

From the experimental results in Table 8, the accuracy of the method proposed in this paper is improved by 0.91\% $\sim$ 15.33\% compared with other methods. This indicates that the method in this paper has a high utilization rate of feature information, and can make full use of the feature information in the data when the sample size is not sufficient.

\section{Conclusions}
In this paper, we propose a method for source cell-phone recognition, which effectively utilizes the short-term spatial information and long-term temporal information in audio data, and provides an in-depth analysis of its spatial complexity and temporal complexity, and achieves good recognition results in source cell-phone recognition research. First, we propose SGMM features, a more comprehensive characterization of the source cell-phone related information. Then, we construct a spatio-temporal information representation learning model utilizing 3D convolutional networks and BiLSTM to sequentially extract the spatial and temporal information of the features, so that the limited data information can be more fully utilized and achieve good experimental results in the source cell-phone recognition.

In the future, we will continue to improve the following two aspects: in feature extraction, we will further explore feature extraction methods with higher density of source cell-phone related information; in recognition models, we will try to investigate how to improve the accuracy of source cell-phone recognition while reducing computational cost. We will also conduct research on recording device recognition for audio data in complex noise scenes, and research more robust source cell-phone recognition methods to achieve a breakthrough in cell-phone recognition technology in complex application scenarios.


 \bibliographystyle{elsarticle} 
 \bibliography{refs}





\end{document}